\documentstyle[12pt,preprint,aps,floats,epsf]{revtex}

\begin{document}
\title{Diffusion of Neutrinos in Proto-Neutron Star Matter with Quarks}
\author{Andrew W. Steiner, Madappa Prakash, and James M. Lattimer}
\address{
Department of Physics \& Astronomy, SUNY at Stony Brook, Stony Brook, NY 
11794--3800}
\date{\today}
\maketitle
\begin{abstract}
Neutrino opacities important in the evolution of a proto-neutron star
containing quark matter are studied.  The results for pure quark
matter are compared with limiting expressions previously derived, and
are generalized to the temperatures, neutrino degeneracies and lepton
contents encountered in a proto-neutron star's evolution.  We find
that the appearance of quarks in baryonic matter drastically reduces
the neutrino opacity for a given entropy, the reduction being
sensitive to the thermodynamic conditions in the mixed quark-hadron
phase.

\end{abstract}


A general picture of the early evolution of a proto--neutron star
(PNS) is becoming well established
\cite{bl,physrep,kj,burrows99,pons,pons2}.  Neutrinos are produced in large
quantities by electron capture as the progenitor star collapses, but
most are temporarily prevented from escaping because their mean free
paths are considerably smaller than the radius of the star. During
this trapped-neutrino era, the entropy per baryon $s$ is about 1
through most of the star and the total number of leptons per baryon
$Y_L = Y_e + Y_{\nu_e} \simeq 0.4$. The neutrinos trapped in the core
strongly inhibit the appearance of exotic matter, whether in the form
of hyperons, a Bose (pion or kaon) condensate or quarks, due to the
large values of the electron chemical potential. As the star cools,
the neutrino mean free path increases, and the neutrinos eventually
leak out of the star, on a timescale of 20-60 s.  During
deleptonization, neutrino diffusion heats the matter to an
approximately uniform entropy per baryon of 2.  If the strongly
interacting components consist only of nucleons, the maximum
supportable mass increases.  In the case that hyperons, a Bose
condensate (pion, kaon) or quarks appear in the core as the neutrinos
leave, the maximum mass decreases with decreasing leptonic content.
Neutron stars which have masses above the maximum mass for completely
deleptonized matter are thus metastable, and will collapse into a
black hole during deleptonization.  Alternatively, if the mass of the
neutron star is sufficiently small, the star remains stable and cools
within a minute or so to temperatures below 1 MeV as the neutrinos
continue to carry energy away from the star.

The way in which this picture is modified when the core of a PNS
contains deconfined quark matter is only beginning to be investigated
\cite{physrep,pcl,carter00,spl}.  In his seminal paper, Iwamoto
\cite{iwam} noted that the non-degenerate $\nu$ mean free path in cold
quark matter is about ten times larger than in nucleonic matter.  We
find that in PNS matter, in which quarks appear towards the end of
deleptonization, similarly large enhancements persist even up to the
largest relevant temperatures ($\sim 30-40$ MeV\cite{spl}), inasmuch
as quarks remain largely degenerate.  On this basis, it can be
anticipated that the presence of quark matter increases the neutrino
fluxes while simultaneously decreasing the deleptonization time,
relative to matter without quarks.  In work to be reported elsewhere,
we explore the possibility that such a change might be detected from a
Galactic supernova in current and planned neutrino detectors. This
would have direct implications for the theoretical understanding of
the high-density regime of QCD which is inaccessible to high energy
Relativistic Heavy-Ion Collider experiments, and, currently, to
lattice QCD calculations at finite baryon density.

To perform detailed simulations of the neutrino signal from a PNS
containing quark matter, as has already been done for matter
containing nucleons, hyperons and/or a kaon condensate
\cite{bl,physrep,kj,burrows99,pons,pons2}, consistent calculations of
neutino interactions in hot lepton-rich matter containing quarks are
required.  It is most likely that quarks exist in a mixed phase with
hadrons \cite{pcl,spl,glen}.  Steiner, et. al.  \cite{spl}, recently
showed that the temperature of an adiabat decreases as a function of
density in a mixed phase of quarks and nucleons. Because $\nu-$cross
sections usually scale with $T^2$, this suggests that the presence of
quark matter might influence the neutrino signal of a PNS with quarks.

In this work, we calculate the diffusion coefficients of neutrinos 
in a mixed phase of hadrons and quarks for the temperatures,
neutrino degeneracies, and lepton contents likely to be encountered in 
the evolution of a PNS with quarks.  We demonstrate that the cross
sections for scattering and absorption of neutrinos by nucleons,
leptons, and quarks are reduced to two integrals, whose integrands are
products of simple polynomials and thermal distribution functions.
The limiting behaviors of the cross sections, for non-degenerate and
degenerate neutrinos, respectively, are compared with previous
calculations \cite{iwam} in the case of pure quark matter.  For
simulations of PNS evolution using the diffusion approximation,
diffusion coefficients, which are energy weighted averages of neutrino
cross sections, are required for matter in which quarks exist in a
mixed phase with hadrons.  We examine the relevant diffusion
coefficients for two thermodynamic conditions especially germane to
PNS evolution.  The first situation is when neutrinos are trapped,
$s\approx1$, and the total lepton content of the matter
$Y_L=Y_e+Y_{\nu_e}$ (which measures the concentrations of the leptons per
baryon) is approximately 0.4.  We also consider the situation when
neutrinos have mostly left the star and the matter has been
diffusively heated ($Y_\nu\approx0, s\approx2$).  We discuss the
impact these results might have upon the evolution of a PNS which
contains quarks in a mixed phase.

For the $\nu-$energies of interest, the neutral
and charged current $\nu-$interactions with the matter in a
PNS are well described by a current-current Lagrangian \cite{weinberg}
\begin{eqnarray}
{\cal L} = \frac {G_F^2}{\sqrt 2} 
\left( \overline\psi_{\nu}(1)
(1-\gamma_{5}) \psi_{p_3}(3) \right) 
\left( \overline\psi_{p_2}(2)
({\cal V} - {\cal A}\gamma_{5}) \psi_{p_4}(4) \right) + 
{\rm {H.C.}} \,,
\end{eqnarray} 
where ${\cal V}$ and ${\cal A}$ are the vector and axial--vector coupling
constants (See Table~1) and $G_F\simeq 1.17$ $\rm{GeV}^{-2}$ is the
Fermi weak coupling constant.  The subscripts $i=1,2,3$, and 4 on the 
four-momenta $p_i$ denote the
incoming neutrino, the incoming lepton, baryon or quark, the outgoing neutrino
(or electron), and the outgoing lepton, baryon or quark, respectively.  The
charged current reactions contribute to absorption of neutrinos by
baryons or quarks and scattering of neutrinos with leptons of the same
generation.  The neutral current interactions contribute to scattering
of neutrinos with leptons and baryons or quarks.  The charged current
contribution to neutrino--lepton scattering in the same generation can
be transformed into the neutral current form, which modifies the
constants ${\cal V}$ and ${\cal A}$ for that case.  For completeness,
the values of ${\cal V}$ and ${\cal A}$ for electron neutrinos are
given in Table 1.  The corresponding values for reactions with
electron anti-neutrinos are obtained by the replacement ${\cal
A}\rightarrow -{\cal A}$.

From Fermi's golden rule, the cross section
per unit volume (or inverse mean free path) is 
\begin{equation}
\frac{\sigma}{V}=g \int \frac{d^3 p_2}{\left(2 \pi\right)^3}
\int \frac{d^3 p_3}{\left(2 \pi\right)^3}
\int \frac{d^3 p_4}{\left(2 \pi\right)^3}~~W_{fi} f_2
\left(1-f_3\right)\left(1-f_4\right)
\left(2 \pi\right)^4 \delta^4\left(p_1+p_2-p_3-p_4\right) \,,
\label{sigma1}
\nonumber
\end{equation}
where the degeneracy factor $g$ is 6 (3 colors $\times$ 2 spins) 
for reactions involving quarks while it is 2 (2 spins) for 
baryons of a single species.    
The Fermi--Dirac distribution functions are denoted by  
$f_i = \left[1+\exp\left(\frac{E_i-\mu_i}{T}\right)\right]^{-1}$, 
where $E_i$ and $\mu_i$ are the energy and chemical potential of
particle $i$.  The transition probability $W_{fi}$,
summed over the inital states and averaged over the final states, is 
\begin{eqnarray}
W_{fi} =  
\frac{G_F^2}{E_1 E_2 E_3 E_4}\left[\left({\cal V}+{\cal A}\right)^2
\left(p_1 \cdot p_2 \right)\left(p_3 \cdot p_4 \right)
+ \left({\cal V}-{\cal A}\right)^2
\left(p_1 \cdot p_4 \right)\left(p_3 \cdot p_2 \right)\right. 
\left. - \left({\cal V}^2-{\cal A}^2\right)
\left(p_1 \cdot p_3 \right)\left(p_4 \cdot p_2 \right)
\right] \,.
\end{eqnarray}
Utilizing $d^3 p_i = p_i^2 ~ d p_i d \Omega_i = p_i E_i
~ d E_i d \Omega_i$ and integrating over $E_4$, Eq.~(\ref{sigma1}) may
be cast in the form
\begin{eqnarray}
\frac{\sigma}{V} & = & \frac{g G_F^2}{32 \pi^5}
\int_{M_2}^{\infty} d E_2 \int_{0}^{\infty} d E_3 ~S ~\frac{E_3}{E_1}
|{\vec p_2}| |{\vec p_4}| 
~\left[
\left({\cal V}+{\cal A}\right)^2 I_{a}+
\left({\cal V}-{\cal A}\right)^2 I_{b}+
\left({\cal V}^2-{\cal A}^2\right) I_{c}
\right] \,,
\label{sigma2}
\end{eqnarray}
where $S=f_2 \left(1-f_3\right)\left(1-f_4\right)$, $M_i$ is the mass of 
particle $i$, and
\begin{eqnarray}
I_{a} = \int d {\Omega}_2 d {\Omega}_3 
d {\Omega}_4~~\delta^3\left(p_1+p_2-p_3-p_4\right) 
\left(p_1 \cdot
p_2\right)\left(p_3 \cdot p_4\right) \,.
\end{eqnarray}
The integrals $I_{b}$ and $I_{c}$ are defined similarly to
$I_{a}$, and all can be performed analytically.  Explicitly, 
\begin{eqnarray}
I_{a}=\frac{\pi^2}{5 p_1 p_2 p_3 p_4}
\left[
3 \left(P_{\rm max}^5-P_{\rm min}^5\right)
-10 \left(A+B\right) \left(P_{\rm max}^3-P_{\rm min}^3\right)
+60 A B \left(P_{\rm max}-P_{\rm min}\right)
\right] \,, 
\label{ia}
\end{eqnarray}
where 
\begin{eqnarray}
2A &=& 2 E_1 E_2+p_1^2+p_2^2,~~2B=2 E_3 E_4+p_3^2+p_4^2, \nonumber \\
P_{\mathrm{min}} &=& \mathrm{max}\left(|p_1-p_2|,|p_3-p_4|\right),~~ 
P_{\mathrm{max}}=\mathrm{min}\left(p_1+p_2,p_3+p_4\right) \,. 
\label{ib}\end{eqnarray}
In the above expression, $p_i \equiv {\vec p_i}$.  $I_{b}$ and $I_c$
are defined to be the same as $I_{a}$, but with appropriate replacements:
\begin{eqnarray}
I_b \equiv I_a(p_2\leftrightarrow p_4, E_2\leftrightarrow -E_4) \,, \qquad
I_c \equiv I_a(p_2\leftrightarrow p_3, E_2\leftrightarrow E_3) \,. 
\label{ic}
\end{eqnarray}
In Eqs.~(\ref{ib}) and (\ref{ic}), $E_4=E_1+E_2-E_3$ and $|{\vec
p_4}|=\sqrt{\left(E_1+E_2-E_3\right)^2-M_4^2}$.  Eqs.~(\ref{sigma2})
through (\ref{ic}) are the principal results of this work and enable
us to compute, for arbitrary conditions of neutrino degeneracy and
matter's temperature, the neutrino diffusion coefficients required in
simulations of PNSs containing quark matter.  We note that similar
techniques were employed to calculate $\nu-$emissivities in cold
catalyzed neutron stars in Refs. \cite{wadhwa,ghosh}.

In limiting cases when the neutrinos are either degenerate or
non-degenerate, and the quarks, which are always degenerate in PNSs,
are massless, simple analytical expressions for the cross section may
be obtained by replacing momenta by Fermi momenta and energies by
chemical potentials in the integrals $I_{a}$, $I_{b}$, and
$I_{c}$. For the sake of comparing such limits with the general
results obtained from Eq.~(\ref{sigma2}), we record various limiting
forms obtained earlier in Refs.~\cite{iwam,tubbs}.
  
\noindent (1) {\it Scattering of degenerate
neutrinos:} The result is the same as for neutrino-electron
scattering:
\begin{eqnarray}
\frac{\sigma_{S}}{V} =  \frac{G_F^2 \mu_2^3}
{5 \pi^3} 
\left[\left(E_1-\mu_1\right)^2+\pi^2 T^2\right] 
\left(\frac{x E_1}{\mu_2}\right)^{1/2}
\left[\left({\cal V}^2 + {\cal A}^2\right)\left(10+x^2\right)+5 \left(
2 {\cal V} {\cal A}
\right) x\right] \,,
\label{ids}
\end{eqnarray}
where $x=\rm{min}(E_1,\mu_2)/\rm{max}(E_1,\mu_2)$.  Here, and in the
following, we have removed the factor
$1-f_1=\left(1+e^{-(E_1-\mu_1)/T}\right)^{-1}$ from \cite{iwam,tubbs}
to obtain transport mean free paths.

\noindent (2) {\it Scattering of non-degenerate neutrinos:}
The inverse scattering mean free path is 
\begin{eqnarray}
\frac{\sigma_{S}}{V} = 
\frac{G_F^2 E_1^3 \mu_2^2}{5 \pi^3} \,.
\label{inds}
\end{eqnarray}
when it is additionally assumed that $\mu_2$ is large compared to $E_1$.  

\noindent (3) 
{\it Absorption of degenerate neutrinos}:
\begin{eqnarray}
\frac{\sigma_{A}}{V}= 
\frac{ 2 G_F^2 \mu_3^3 }{ 5 \pi^3 \mu_1^2}
\left(10 \mu_4^2 + 5 \mu_4 \mu_3 + \mu_4^2 \right)
\left[\left(E_1-\mu_1\right)^2+\pi^2 T^2\right] \,.
\label{adn}
\end{eqnarray}

\noindent (4) {\it Absorption of non-degenerate neutrinos}: 
The general result is greatly simplified by additionally  
assuming that the quark chemical potentials are modified by perturbative
gluon exchange:
\begin{eqnarray}
\frac{\sigma_{A}}{V}= 
\frac{16}{\pi^4}
\alpha_c G_F^2 p_{F_2} p_{F_3} p_{F_4}
\left[E_1^2+\pi^2 T^2\right] \,.
\label{and}
\end{eqnarray}

Representative cross sections from Eq.~(\ref{sigma2}) are compared
with the limiting forms in Eqs.~(\ref{ids})--(\ref{and}) in
Fig.~\ref{iw4}.  Degenerate neutrinos are assumed in this figure to
have $\mu_\nu>>T$ and non-degenerate neutrinos are assumed to have
$\mu_\nu\approx0$.  In the regions where they were expected to be
valid, namely $E_\nu/T>>1$ for degenerate absorption and scattering,
and also non-degenerate scattering, and $E_\nu\approx T$ for
non-degenerate absorption, the limiting forms give adequate
representations of the general results.  However, significant
deviations occur in the cases of non-degenerate absorption when $E_\nu
\ne T$, and for degenerate absorption and non-degenerate scattering
when $E_\nu<<T$.  The deviation for non-degenerate absorption is due
to the neglect of $p_\nu$ in the momentum conservation condition in
Eq.~(\ref{sigma1}) in Ref. \cite{iwam} (which is appropriate for cold
catalyzed stars, but not for hot matter in PNSs), which limits its
applicability to the region $E_\nu \approx T$.  The other two
deviations are simply due to the assumption in Ref. \cite{iwam} that
$E_\nu>>T$.

The weak interaction timscales of neutrinos are much smaller than the
dynamical timescale of PNS evolution, which is on the order of
seconds.  Thus, until neutrinos enter the semi-transparent region,
they remain close to thermal equilibrium in matter. Hence, neutrino
propagation may be treated in the diffusion approximation with the
differential equations for the flux of energy ($F_{\nu}$) and lepton
number ($H_{\nu}$) \cite{pons}:
\begin{eqnarray}
H_{\nu} & = & -\frac{T^2 e^{-\Lambda-\phi}}{6 \pi^2}\left[D_4 
\frac{\partial \left(T e^{\phi}\right)}{\partial r} +
\left(T e^{\phi}\right) 
D_3 \frac{\partial \eta(r)}{\partial r}\right] \qquad {\rm and} \qquad 
\nonumber \\
F_{\nu} & = & -\frac{ T^3 e^{-\Lambda-\phi}}{6 \pi^2}\left[D_3 
\frac{\partial \left(T e^{\phi}\right) }{\partial r} +
\left(T e^{\phi}\right) 
D_2 \frac{\partial \eta(r)}{\partial r}\right] , \nonumber 
\end{eqnarray}
where $\Lambda$ and $\phi$ are general relativistic metric functions,  
$\eta=\mu_{\nu}/T$
and $D_2$, $D_3$, and $D_4$ are diffusion coefficients decomposed as 
\begin{eqnarray}
D_2 = D_2^{\nu_e} + D_2^{\bar{\nu}_e} , \qquad
D_3 = D_3^{\nu_e} - D_3^{\bar{\nu}_e} , \qquad
D_4 = D_4^{\nu_e} + D_4^{\bar{\nu}_e} + 4 D_4^{\nu_{\mu}} \,. 
\end{eqnarray}
The transport of $\mu$ and $\tau$ neutrinos and anti-neutrinos are
well approximated \cite{bl} by assuming that they contribute equally to
$D_4$ and are represented by $D^{\nu_{\mu}}$.
These coefficients are defined in terms of the energy
dependent diffusion coefficient $D^p(E_{\nu})$ by
\begin{equation}
D_n^p=\int_0^{\infty} dx ~x^n D^p(E_1) f (E_1) \left[1-f(E_1)\right] \,,
\end{equation}
where $x=E_1/T$, and the superscipt 
$p$ denotes either the electron neutrino, the
anti-electron neutrino, or the $\mu$ and $\tau$ neutrinos and thier
antineutrinos.  In turn, the energy dependent diffusion coefficient is
obtained directly from the cross sections per unit volume through
\begin{equation}
\left(D^p(E_1)\right)^{-1} = \frac{1}{1-f_1}
\left[
\sum_{\rm{r=(p,L)}} \frac{\sigma_r}{V} +
\chi \sum_{\rm{r=(p,H)}} \frac{\sigma_r}{V} +
\left(1-\chi\right) \sum_{\rm{r=(p,Q)}} \frac{\sigma_r}{V} 
\right]  \,,
\end{equation}
where the $(p,L)$, $(p,H)$, and $(p,Q)$ represent the sum over all the
reactions of particle $p$ with leptons, hadrons, or quarks,
respectively. The factor $\left(1-f_1 \right)^{-1}$ ensures detailed
balance, and $\chi$ is the volume fraction of matter in the hadronic
phase.

Note that in the case of scattering, the Pauli blocking factor
corresponding to the outgoing neutrino is omitted, since the neutrino
distribution function is not always known a priori unless a full
transport scheme is employed.  It is possible, however, to devise a
simplified scheme \cite{pons} in which the dependence on the neutrino
distribution function is minimized. Such a scheme is valid only when
scattering from light particles (either electrons or quarks) does not
dominate the opacity. Our results below show that this requirement is
indeed met, because absorption dominates over scattering by a factor
of 2 to 5 at all densities interior to the central densities of PNSs
investigated here.

We describe neutron star matter at finite density and temperature 
using the Gibbs phase rules
\cite{spl,glen}. The conditions of baryon number density and charge
conservation in the mixed phase are
\begin{eqnarray*}
n_B = \chi n_B^H + \left(1-\chi\right) n_B^Q , \qquad 
0 = \chi n_c^H + \left(1-\chi\right) n_c^Q + n_c^L \,,
\end{eqnarray*}
where $n_B$ and $n_c$ are the baryon number and charge densities,
respectively; $H$, $Q$, and $L$ denote 
hadrons, quarks, and leptons. 
Since the dynamical time scale is much longer
than the weak interaction time scale, beta equilibrium 
implies that the various
chemical potentials satisfy the relations
\begin{eqnarray}
\mu_e-\mu_{\nu_e} = \mu_\mu - \mu_{\nu_\mu}\,;\qquad \mu_B = b_i \mu_n
-q_i (\mu_e - \mu_{\nu_e}) \,,
\end{eqnarray}
where $b_i$ and $q_i$ are the baryon number and electric charge of the
hadron or quark of species $i$.  When the neutrinos are trapped, the
electron lepton number $Y_{L} = (n_e + n_{\nu_e})/n_B$ is initially
fixed at a value $\simeq 0.4$ as suggested by collapse calculations.

Hadronic matter is described using a field-theoretical description in
which nucleons interact via the exchange of $\sigma$, $\omega$, and
$\rho$ mesons. The meson-nucleon couplings and the couplings of the
$\sigma$ self-interaction terms are determined by reproducing the
empirical properties of nuclear matter, $E_B=-16.0$ MeV,
$M^{*}/M=0.6$, $K=250$ MeV, $a_{sym}=32.5$ MeV, and
$n_0=0.16~\rm{fm}^{-3}$.  Quark matter is described using the MIT Bag
model, with a bag constant of $B=200~\rm{MeV}/\rm{fm}^3$.  (Similar
results are obtained with four-quark interactions in the
Nambu--Jona-Lasinio model \cite{spl}.) The mixed phase is assumed to
be homogeneous.   For more details of the
calculation of the EOS, see Ref. \cite{spl}.



The cross sections per unit volume (or inverse mean free paths) for
$\nu_e$ scattering and absorption are shown in Fig. \ref{cs3} for the
two stages of PNS evolution described before.  It is important to
recall that these curves are drawn under conditions of fixed entropy.
(Constant entropy adiabats shown in Fig.~3 of Ref.~\cite{spl} are
helpful to gain insights into the behavior of the cross sections shown
here.)  The individual contributions from the different reactions in
the pure nucleon and quark phases (thin lines) and in the mixed phase
(thick lines) are marked in this figure.  The vertical dashed lines
show the central densities of $1.4$M$_{\odot}$ and the maximum mass
configurations, respectively.  Notice that quarks exist only in the
mixed phase; the pure quark phase occurs at densities above the
central densities of maximum mass stars in all cases shown here.

In general, for a given density and temperature, the pure quark-phase
opacity (or equivalently the cross section per unit volume) is less
than that of hadrons due to the former's smaller matrix elements
sampled by a relativistic phase space.  In addition, for a given
entropy and density, pure quark matter favors a lower temperature than
hadronic matter \cite{spl}.  It is natural, therefore, that within the
mixed phase region of hadrons and quarks, the net cross section per
unit volume either flattens or decreases with increasing density.  The
reduction of opacities from that of the pure hadronic phase is
enhanced in the $\nu-$free case, reflecting the more extreme decrease
of temperature across the mixed phase region in that case \cite{spl}.
The precise density dependence of the net neutrino opacity depends
upon the details of the mixed phase.

Note that the total absorption cross section is larger than the
scattering cross section for both degenerate and nondegenerate
situations.  As discussed above, this justifies our approximate
treatment of scattering in the calculation of the diffusion
coefficients.

The diffusion coefficients most relevant for the PNS simulations, in
matter with and without a mixed phase of hadrons and quarks, are $D_2$
and $D_4$, and these are shown in Fig. \ref{dc3}.  
Insofar as absorption dominates scattering, the behavior of
these coefficients can be understood qualitatively by utilizing the limiting 
forms for neutrino absorption in the degenerate (Eq.~({\ref{adn})) and 
nondegenerate (Eq.~(\ref{and})) cases, respectively. 
The actual behavior is somewhat more complicated, but this
assumption will suffice for a qualitative interpretation of
Fig. \ref{dc3}.  In this case, the leading behaviors may be extracted 
to be 
\begin{eqnarray}
D_2 \propto \lambda(\mu_\nu/T) (\mu_\nu/T)^2 \,\qquad  {\rm and} \qquad
D_4 \propto \lambda(\mu_\nu/T=0)\,,
\end{eqnarray}
where the mean free path $\lambda=(\sigma/V)^{-1}$.  In these equations, 
$D_2$ is evaluated under conditions of extreme neutrino degeneracy and
$D_4$ is evaluated assuming that $\mu_\nu=0$. Thus, both $D_2$ and
$D_4$ should simply reflect the inverse behavior of the cross section
per unit volume, which decreases with increasing density in the pure
phases, but increases within the mixed phase region.

Concerning the evolution of a PNS, we expect that the initial star,
which is lepton rich, will not have an extensive mixed phase region.
Only after several seconds of evolution will quark matter appear.  In
the newly-formed mixed phase region, the neutrino opacity will be
substantially smaller than in the case in which a mixed phase region
does not appear.  However, due to the large $\nu-$optical depth of the
PNS, neutrinos remain trapped, and no significant effect on emergent
neutrino luminosities is expected at early times.  As the star
evolves, however, the relatively larger increase in opacity (note the
increases in $D_4$ relative to $D_2$) and the growing extent of the
mixed phase region eventually allows a larger flux of neutrinos, and
thereby a more rapid evolution.

In summary, we have calculated neutrino opacities for matter
containing quarks for the temperatures, neutrino degeneracies and
lepton contents relevant for PNS simulations, employing Gibbs phase
rules to construct a mixed hadron-quark phase.  We find that, in the
presence of quarks, neutrinos have a significantly smaller opacity and
hence larger diffusion coefficients than those in purely hadronic
matter at similar densities.  These differences may have an observable
impact on the neutrino flux from PNSs containing quark matter, but
these differences are not expected to become apparent until the PNS is
10--20 s old.  Simulations of PNSs with a mixed phase of hadrons and
quarks are under investigation \cite{Pons01}.  The influence of
heterogeneous structures \cite{reddy00,christiansen00} and
superfluidity \cite{carter00} in matter will be addressed in future
work.


We acknowledge research support from the U.S. Department of Energy
under contracts DOE/DE-FG02-88ER-40388 (AWS and MP) and
DOE/DE-FG02-87ER-40317 (JML). We thank Jose A. Pons and Sanjay Reddy
for useful discussions.

\newpage

\begin{table}
\begin{center}
\noindent{TABLE 1: The standard model charged and neutral current
vector and axial--vector couplings of neutrinos to leptons, baryons,
and quarks; $\theta_C$ is the Cabbibo angle ($\cos{\theta_C}=0.973$),
$\theta_W$ is the weak mixing angle ($\sin{\theta_W}=0.231$), and
$g_A=1.23$ is the baryon axial--vector coupling constant.  }
\begin{tabular}{cccc}
& $1+2\rightarrow 3+4$ & ${\cal V}$ & ${\cal A}$ \\
\hline
& $\nu_e + \mu^{-} \rightarrow \nu_{\mu} + e^{-}$ & $1$ & $1$ \\
& $\nu_l + n \rightarrow l^{-} + p$ & $\frac{1}{2}\cos{\theta_C}$ & 
$\frac{1}{2}g_A \cos{\theta_C}$ \\
Charged & $\nu_l + d \rightarrow l^{-} + u$ & $\cos{\theta_C}$ & 
$\cos{\theta_C}$ \\
current & $\nu_l + s \rightarrow l^{-} + u$ & $\sin{\theta_C}$ & 
$\sin{\theta_C}$ \\
\hline
& $\nu_e + e^{-} \rightarrow \nu_e + e^{-}$ &$\frac{1}{2}+2\sin^2{\theta_W}$ & 
$\frac{1}{2}$ \\  
& $\nu_e + \mu^{-} \rightarrow \nu_e + \mu^{-}$ & 
$-\frac{1}{2}+2\sin^2{\theta_W}$
& $\frac{1}{2}$ \\  
& $\nu_e + n \rightarrow \nu_e + n $ & $-\frac{1}{2}$ & 
$-\frac{1}{2}g_A$ \\  
Neutral & $\nu_e + p \rightarrow \nu_e + p $ & 
$\frac{1}{2}+2\sin^2{\theta_W}$ & 
$\frac{1}{2}g_A$ \\  
current & $\nu_e + u \rightarrow \nu_e + u $ & 
$\frac{1}{2}-\frac{4}{3}\sin^2{\theta_W}$ & $\frac{1}{2}$ \\  
& $\nu_e + d \rightarrow \nu_e + d $ & 
$\frac{1}{2}+\frac{2}{3}\sin^2{\theta_W}$ & 
$-\frac{1}{2}$ \\  
& $\nu_e + s \rightarrow \nu_e + s $ & 
$\frac{1}{2}+\frac{2}{3}\sin^2{\theta_W}$ & $-\frac{1}{2}$ \\  
\end{tabular}
\label{tab1}
\end{center}
\end{table}

\newpage

\bibliography{spl2}
\bibliographystyle{unsrt}

\newpage
\section*{Figure Captions}
\noindent Figure~\ref{iw4}: $\nu_{e}$ cross sections per unit
volume in pure quark matter, for $T=5$ MeV. Solid lines show complete
results from Eq.~(\ref {sigma2}), and dashed lines indicate limiting
forms from \cite{iwam}. The labels ``Degenerate'' and
``Non-Degenerate'' refer to neutrinos.  Upper panels show results for
two times nuclear matter density, and the lower panels show results
for a neutrino energy of 50 MeV ($\mu$ is the chemical potential of
the incoming quark).  \\

\noindent Figure~\ref{cs3}: $\nu_{e}$ cross sections with
various particles in matter containing a mixed phase of quarks and
hadrons ($n_0=0.16~\mathrm{fm}^{-3}$).  The left panels show
scattering cross sections for neutrinos with the indicated incoming
hadrons, quarks, or leptons. Thick lines show the extent of the mixed
phase region.  The right panels show absorption cross sections on
nucleons and quarks.  The upper panels correspond to the
neutrino-trapped era when $s=1$ and $Y_{L}=0.4$, and the lower panels
to the time following deleptonization when $s=2$ and $Y_{\nu}=0$. The 
vertical dashed lines labelled $u_{1.4}$ and $u_{max}$ indicate the central
densities of a $M_G = 1.4$ M$_{\odot}$ star and the maximum mass star
($M_G=2.22$ M$_{\odot}$ for the upper panels and $M_G=1.89$ M$_{\odot}$
for the lower panels), respectively. \\

\noindent Figure~\ref{dc3}: Diffusion coefficients for the
neutrino-trapped era (left panel) and hot deleptonized era (right
panel). Thick lines show the extent of the mixed phase. Solid lines
correspond to matter with a mixed phase, and dashed lines to matter
containing only nucleons. The vertical dashed lines have the same meaning 
as in Fig. 2.

\newpage

\begin{figure}
\begin{center}
\leavevmode
\setlength\epsfxsize{6.0in}
\setlength\epsfysize{6.0in}
\epsfbox{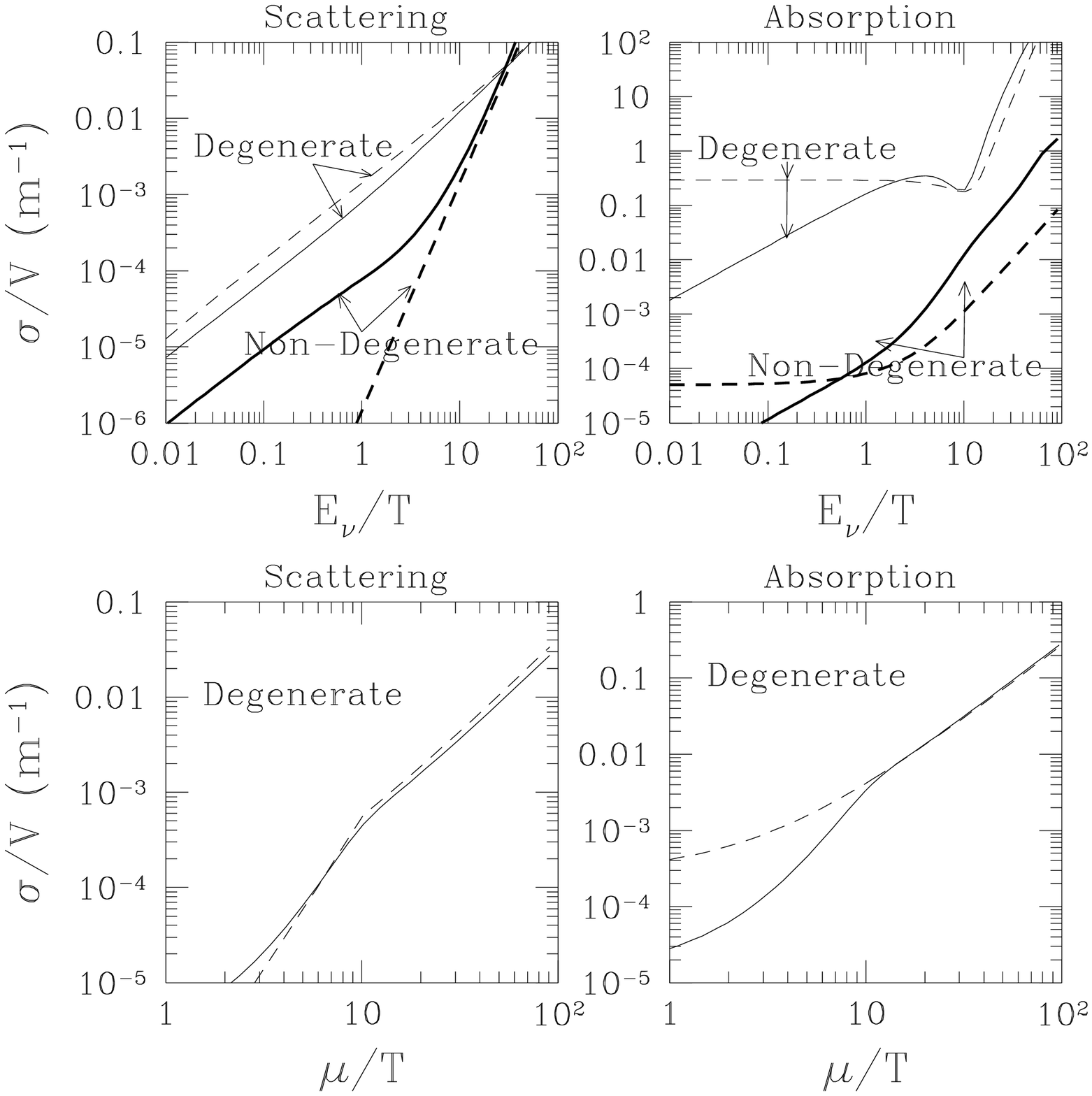}
\caption[]{}
\label{iw4}
\end{center}
\end{figure}

\begin{figure}
\begin{center}
\leavevmode
\setlength\epsfxsize{6.0in}
\setlength\epsfysize{6.0in}
\epsfbox{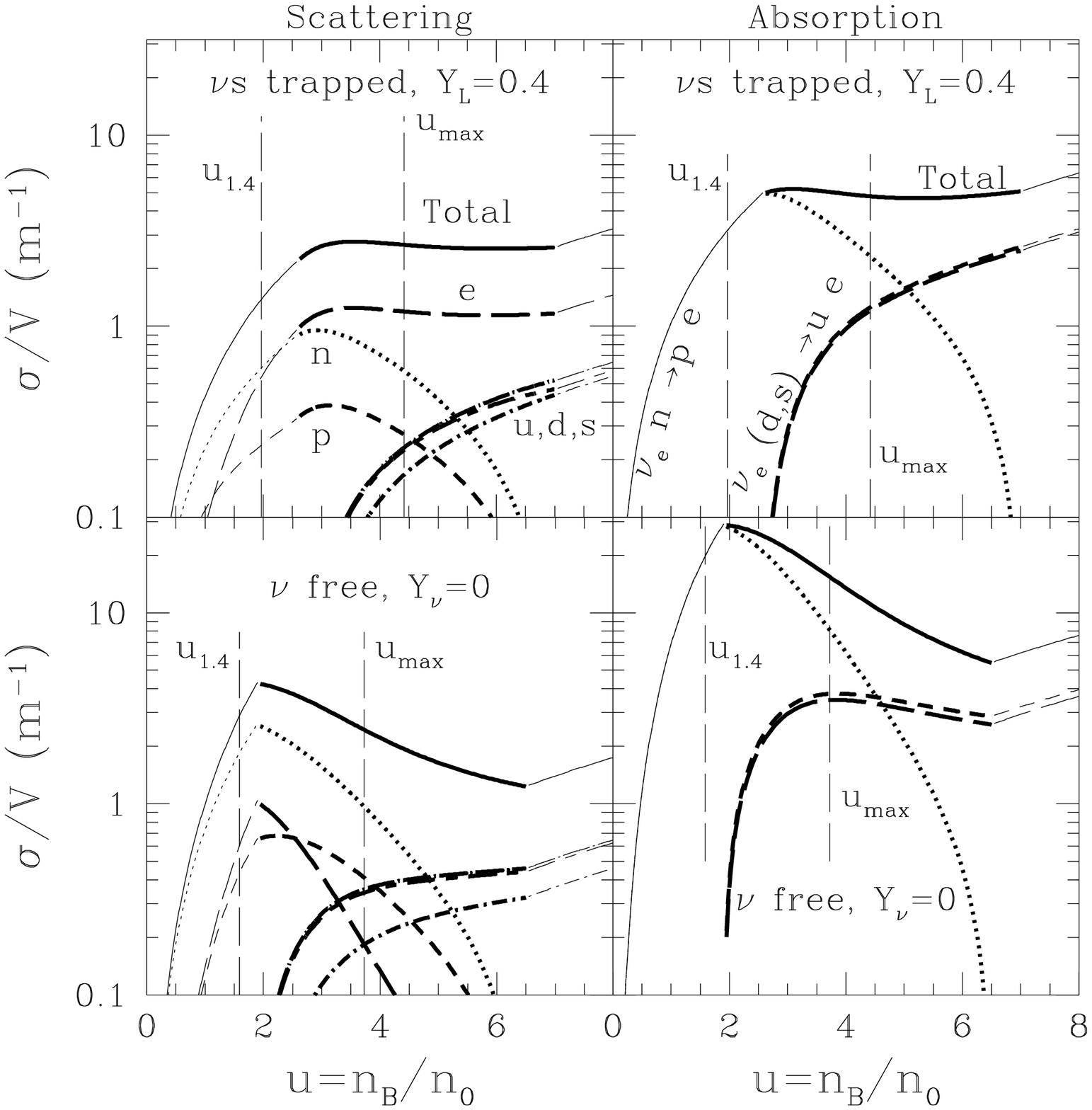}
\caption[]{}
\label{cs3}
\end{center}
\end{figure}

\begin{figure}
\begin{center}
\leavevmode
\setlength\epsfxsize{6.0in}
\setlength\epsfysize{6.0in}
\epsfbox{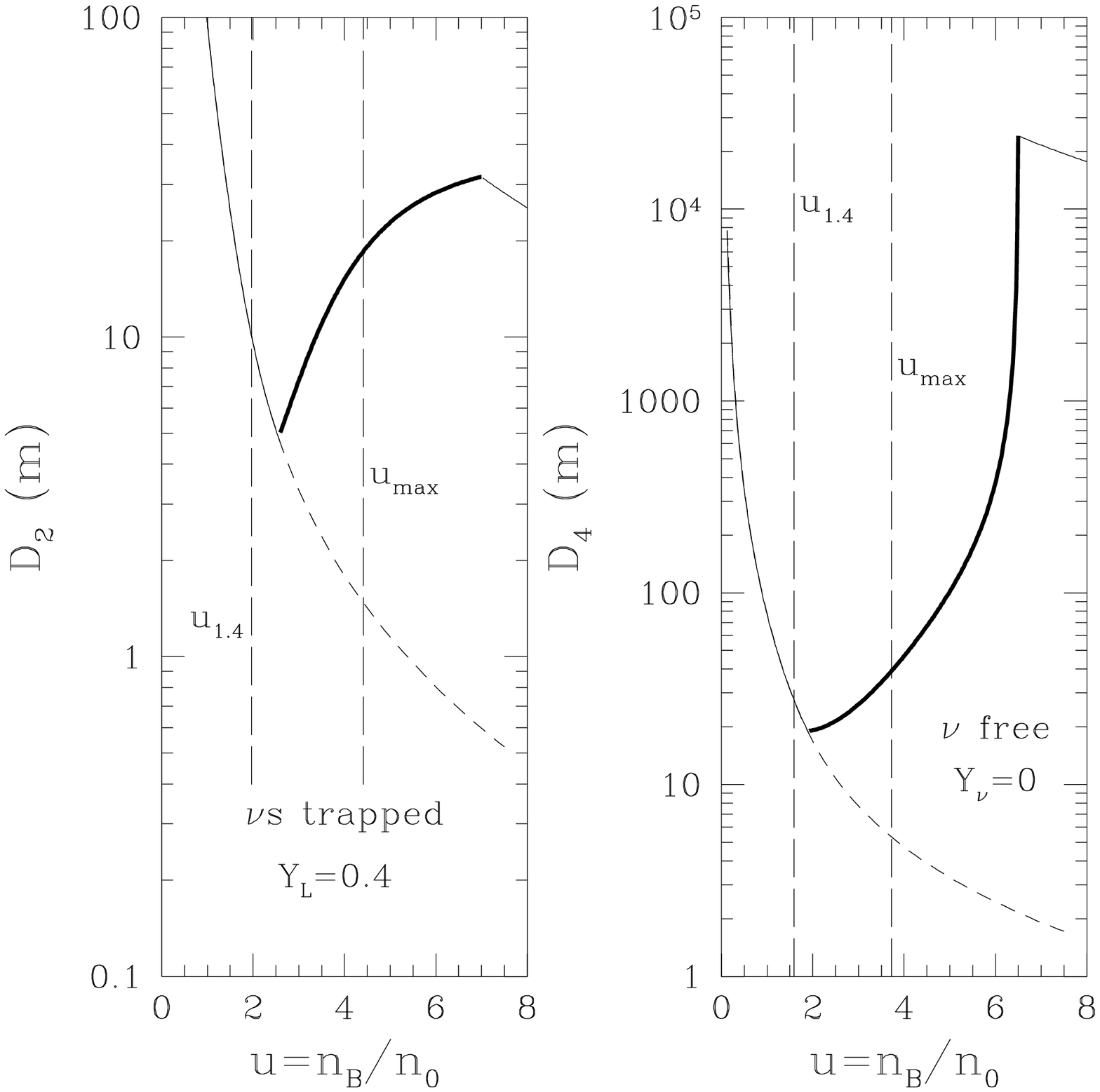}
\caption[]{}
\label{dc3}
\end{center}
\end{figure}

\end{document}